\documentclass[aps,prb,preprint,groupedaddress,showpacs]{revtex4}

\usepackage{graphicx}
\usepackage{dcolumn}
\usepackage{bm}

\begin{document}


\title{Frequency quenching of microwave induced resistance oscillations in a high mobility 2DEG}


\author{S. A. Studenikin}
\email[]{sergei.studenikin@nrc.ca}
\affiliation{Institute for Microstructural Sciences, National Research
Council, Ottawa, Ontario K1A-0R6, Canada}

\author{O.M. Fedorych}
\author{M. Byszewski}
\author{D. K. Maude}
\author{M. Potemski}
\affiliation{High Magnetic Field Laboratory, MPI/FKF and CNRS, Grenoble 38-042, France}

\author{M. Hilke}
\affiliation{Department of Physics, McGill University, Montreal H3A-2T8, Canada}

\author{K. W. West}
\author{L. N. Pfeiffer}
\affiliation{Laboratories, Lucent Technologies, Murray Hill, New Jersey 07974-0636, USA}

\author{A. S. Sachrajda}
\author{J. A. Gupta}
\author{Z. R. Wasilewski}
\affiliation{Institute for Microstructural Sciences, National Research
Council, Ottawa, Ontario K1A-0R6, Canada}


\date{\today}

\begin{abstract}
The frequency dependence of microwave induced resistance oscillations (MIROs) has been studied experimentally in high-mobility electron GaAs/AlGaAs structures to explore the limits at which these oscillations can be observed.   It is found that in \emph{dc} transport experiments at frequencies above ~120 GHz MIROs start to quench while above ~230 GHz they completely disappear.   The results will need to be understood theoretically but are qualitatively discussed within a model in which forced electronic charge oscillations (plasmons) play an intermediate role in the interaction process between the radiation and the single particle electron excitations between Landau levels.
\end{abstract}


\maketitle

\section{Introduction}
The recently observed new phenomenon of microwave induced resistance oscillations (MIRO) on very high mobility samples has attracted considerable attention from the scientific community, partly due the fact that under certain conditions MIRO can evolve into zero-resistance states.\cite{zudov,mani,zudov2,stud1,dorozhkin1}  A number of theoretical models have been suggested to explain this phenomenon based on scattering due to disorder \cite{durst,ryzhii,vavilov,inarrea}, a non-equilibrium distribution function \cite{dmitriev1,dmitriev2}, and a plasmon mechanisms \cite{mikhailov,volkov}.  It should be noted that long before the experimental observation of MIRO, the possibility of such oscillations in resistance and absolute negative conductance were discussed theoretically by Gladun and Ryzhii in 1970 within a scattering model \cite{ryzhii2}.  In ref. \cite{magarill} MIROs were independently predicted in dynamic conductivity within a purely classical approach based on the Boltzmann kinetic equation.  In this paper a 2DEG was subjected to a microwave field and a lateral superlattice potential, which in the context of the MIRO mechanism can be considered as a Fourier harmonic of the scattering potential.  Most of the approaches successfully predict the period and phase of the oscillations.  In the first order approximation they arrive at similar final equations which describe quite well the MIRO waveform in the quasi-classical regime of large filling factors. \cite{stud2}  Nonetheless, in spite of substantial theoretical effort, none of the existing theories can explain other essential MIRO characteristics, including: (i) the independence of the MIRO amplitude on the left/right circular polarization \cite{smet}, (ii) the necessity to assume equal transition probabilities induced by microwaves for electrons between Landau levels (LL)\cite{stud2}, (iii) the large nonlinear effects at fractional harmonics \cite{zudov3}, (iv) why MIRO are only observed in very high mobility samples while the theories need to introduce disorder (dirtier samples) to explain MIRO, (v) why MIRO are not observed in reflection experiments \cite{stud3}, (vi) why MIRO were never reported in the optical or even sub-millimeter frequency range \cite{stud4}, (vii) why magneto-plasmons are very much suppressed in transport experiments even though electrodynamics clearly plays an important role in these experiments. \cite{volkov,mikhailov2}

  In this work we experimentally study the evolution of MIRO over a wide range of frequencies and examine   the importance of electrodynamics by studying microwave (MW) magneto-absorption in 2DEG long rectangular samples.  We find that below 120 GHz the MIRO waveform is described well by the existing models.  At higher frequencies the waveform starts to deviate from the theory and oscillations completely vanish at frequencies beyond 230 GHz.  It should be stressed that not only the amplitude but also the exponential decay factor (equivalent to quantum scattering time) diminishes at higher frequencies.  The results will hopefully motivate theoretical investigations,  but may be qualitatively explained within a model where plasmons (forced charge oscillations) intermediate the interaction process between the microwave radiation and the conduction electrons.

\section{Experimental}

 The experiment was performed on a GaAs/AlGaAs hetero-structure containing a high mobility two dimensional electron gas (2DEG) confined at the hetero-interface.  The size of the cleaved sample was about 2$\times$5 mm$^2$.  After a brief illumination with a red LED, the 2DEG attained an electron concentration of 2.0$\times$10$^{11}$cm$^{-2}$ and a mobility of  8$\times$10$^6$ cm$^2$/Vs at temperature 2 K.   The sample was placed in a $^4$He cryostat equipped with a superconducting solenoid. A tunable klystron MW generator (model  $\Gamma$C-03) in combination with a frequency doubler was used as the microwave source over 80 to 226 GHz range.   The output power was adjusted by a MW attenuator and was kept approximately constant in the 10$^{-3}$ Watt range.  The microwaves were delivered to the sample using a three meter long, 12 mm diameter stainless steel pipe.   The MW power intensity at the sample was in the range of a 100 $\mu$W/cm$^2$. It's relative value was monitored with roughly 50$\% $ accuracy by measuring an Alan Bradley carbon thermo-resistor located near the sample.
Figure 1 (a) shows experimental traces of MIRO for different frequencies between 80 and 226 GHz at T=1.8K.    Two main effects should be emphasized in Fig. 1 (a).  First, the overall amplitude of the oscillations is reduced with increasing MW frequency.  Although, such an amplitude reduction is qualitatively captured by available theories, e.g. see eq. (10) in ref. \onlinecite{dmitriev2} which for $\omega_c\ll \omega$ predicts a $1/\omega^4 $ decrease with increasing microwave frequency, it is difficult and unreliable to make quantitative comparison with the experiment for the following reasons. The available theories are derived for infinite size samples assuming that the MW filed is uniform in the 2DEG plane.
In reality, the microwave field is not very well defined inside the sample due to the high metallic-like reflection by the 2DEG  and finite sample size, which makes it difficult to precisely estimate the spatial filed distribution inside the sample.  Therefore, it is not exactly correct to \emph{a priori} assume that the MW  field is uniform in the 2DEG plane.  The field distribution inside the sample is definitely affected by plasmon oscillations. \cite{mikhailov2}  The importance of the plasmon oscillations in MIROs is still to be understood.

\begin{figure}[h] 
\includegraphics[width=8cm]{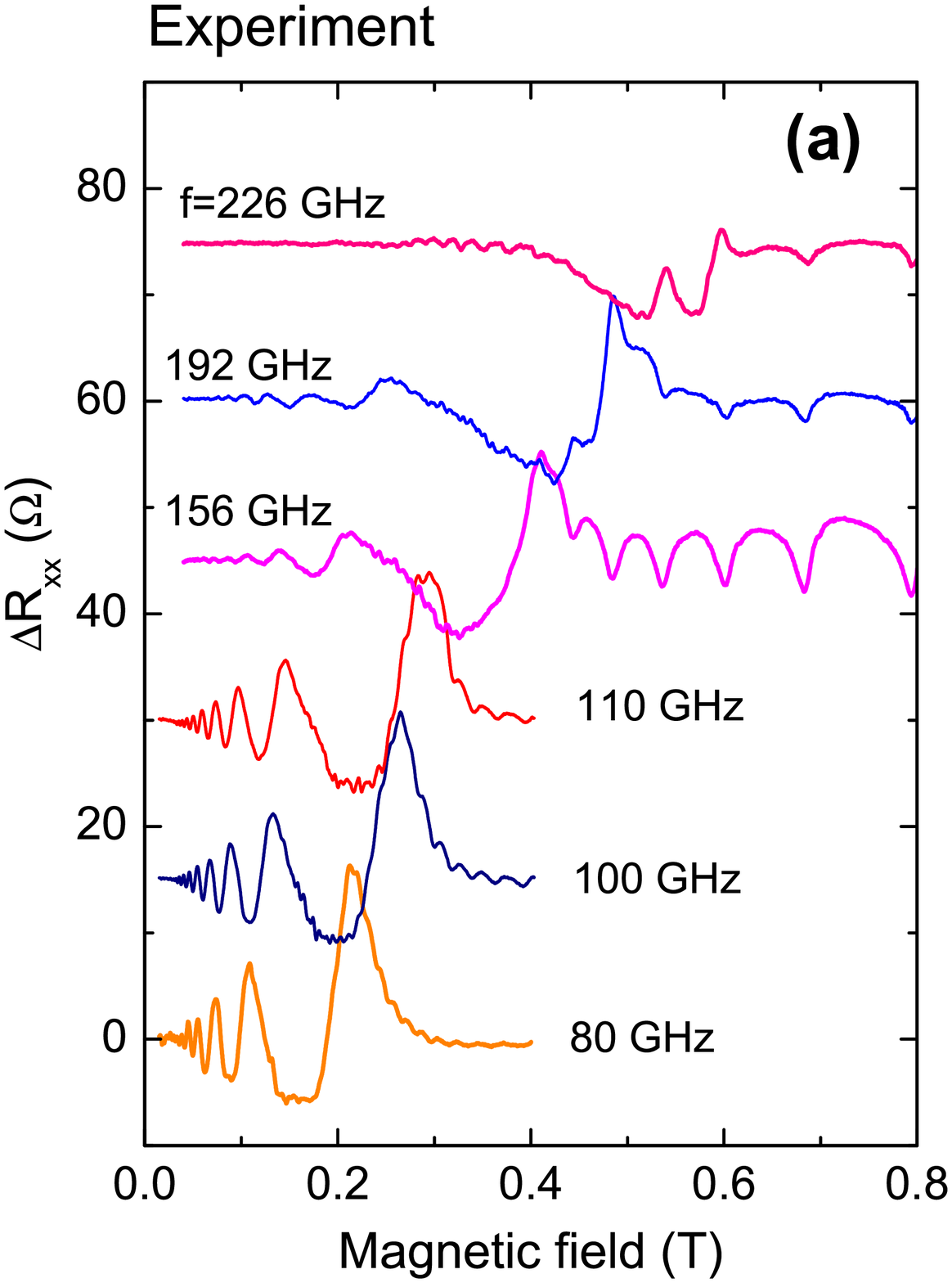}
\includegraphics[width=8cm]{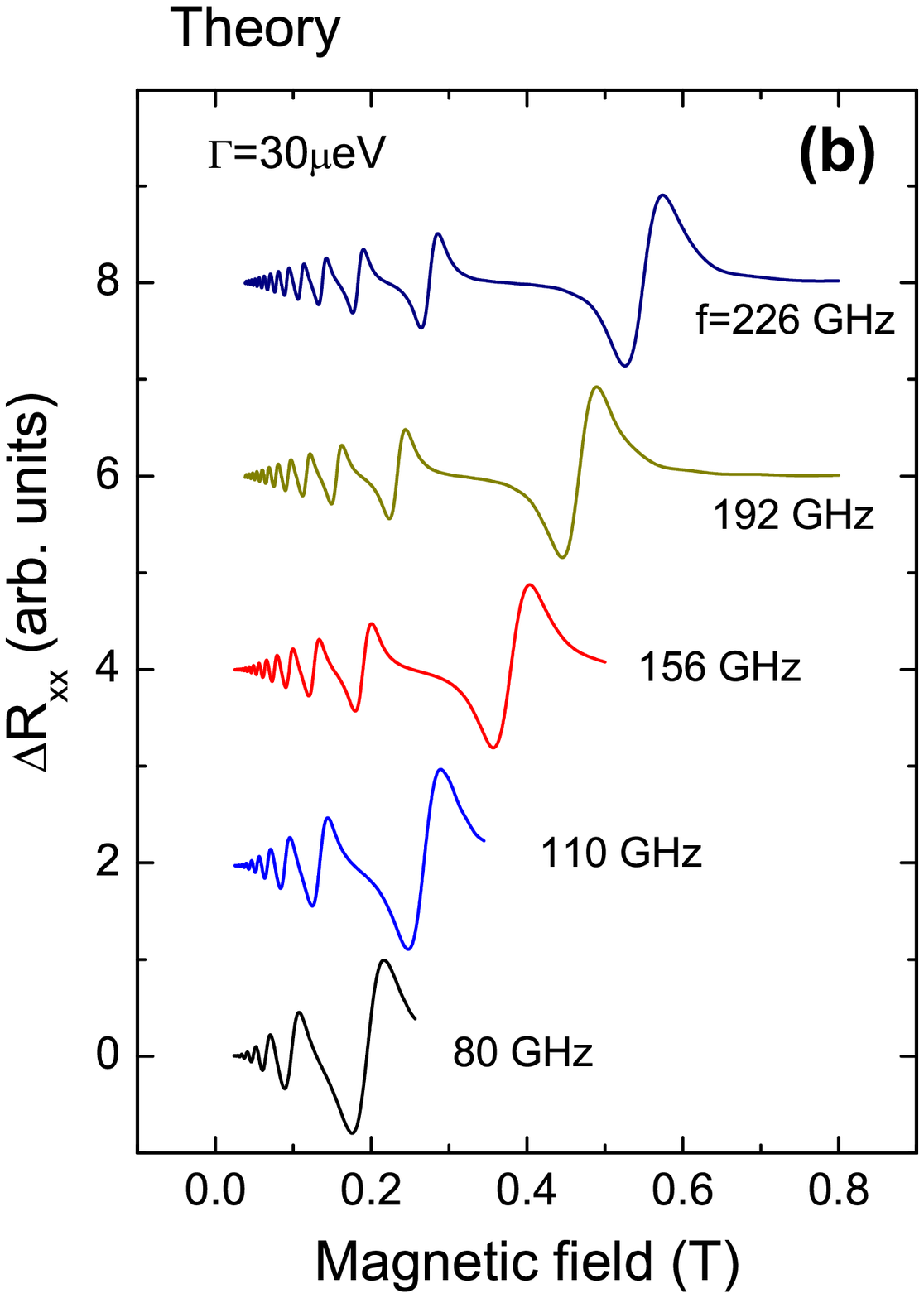}
\caption{(Color online) MIRO ( $\Delta R_{xx}$=$R_{xx}$(MWs on)-$R_{xx}$(no MWs)) for different MW frequencies from 80 to 226 GHz: \textbf{(a)} experimental traces at T=1.8K, \textbf{(b)} corresponding normalized theoretical traces by eq. (1) with the same Landau level width  $\Gamma$=30 $\mu$eV.  Traces are shifted vertically for clarity. }
\end{figure}

The second, more important effect in Fig. 1(a) is a smaller number of observable oscillations for higher frequencies, which is  equivalent to a decrease in the quantum relaxation time, $\tau_q$, that cannot be even qualitatively explained by existing theories.
To explore the last effect in more details, in Fig. 1 (b) we plot normalized theoretical curves corresponding to the experiment in Fig. 1 (a).  The theoretical curves were calculated for a constant Landau level width  $\Gamma$=30 $\mu$eV using the following equation\cite{stud2}:

   \begin{equation}\label{1}
\Delta R_{xx}(B) \propto A\int d\varepsilon \left[ n_{F}(\varepsilon)-n_{F}(\varepsilon +\hbar\omega)\right]\nu (\varepsilon ) \partial_{\varepsilon} \nu (\varepsilon
+\hbar\omega ),
\end{equation}

where  $\nu (\varepsilon )= \sum(eB/\pi^2\hbar\Gamma)/\{1+(\varepsilon-E_i)^2/\Gamma^2\}$
is the electron density of states in a quantizing magnetic field, $\varepsilon$  is electron energy in conduction band, $\omega =2\pi f$ is the cyclic microwave frequency,
$n_F=1/\{1+\exp[(\varepsilon-\varepsilon_F)/k_BT]\} $ is the Fermi distribution function,
$E_i=(i+1/2)\hbar\omega_c $    is the Landau Levels (LL) energy spectrum with i=0,1,2\ldots,   $\omega_c=eB/m^*$ is the cyclotron frequency, and $A(T,\omega )$ is a pre-factor depending on temperature and MW frequency but not magnetic field.  The frequency-dependent pre-factor does not affect the waveform shape vs. magnetic field and therefore can be used as a constant arbitrary parameter when extracting the MIRO decay factor (equivalently Landau level width $\Gamma$ )   from experimental curves vs. magnetic field.
A comparison between Figs. 1 (a) and (b) reveals that the simple equation works relatively well up to 110 GHz while for higher frequencies the experimental waveforms essentially distort and their amplitudes are monotonically reduced until they completely quench at frequencies higher than 226 GHz, while the theory predicts a growing number of oscillations.  In ref. \onlinecite{stud4}  it was shown that at even higher frequencies in the THz range the only remaining feature related to the cyclotron resonance absorption at $\omega = \omega_c$ occurs in the photo-response  $\Delta R_{xx}$ due to a bolometric type of effect and no features related to MIRO can be observed in this frequency range.

\begin{figure}[h]
\includegraphics[width=8cm]{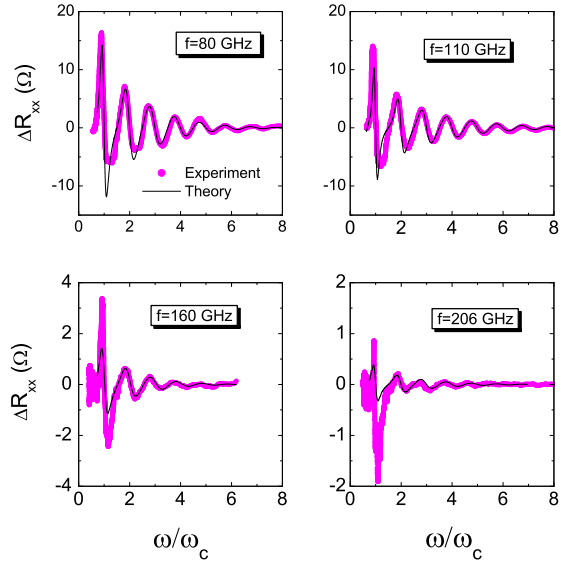}
\caption{(Color on line) Experimental traces of MIRO for different frequencies plotted as a function of    $\omega/\omega_c $ (points) fitted with eq. (1) - solid lines.  }
\end{figure}

Let us examine in more details the parameter $\Gamma$  as a function of the microwave frequency.  Figure 2 shows several MIRO traces as a function of  $\omega/\omega_c$  ($\propto$ inverse magnetic field) for different microwave frequencies.  The experimental traces (points) are fitted by eq.(1) (the solid lines).   Although in the experiment the first harmonics (n=0 and 1) start to deviate from the theory for frequencies above ~120 GHz, it is still possible to fit the higher harmonic oscillations and quantitatively estimate the parameter $\Gamma $  from the fits.  This is plotted in Fig. 3 as a function of the MW frequency.  Since the fits were performed over approximately the same magnetic field range from ~0.02 to 0.4 T, the large changes in $\Gamma$  cannot be attributed to its magnetic field dependence.  From Fig. 3 it is evident that for MW frequencies below 120 GHz $\Gamma$  remains constant, but it increases sharply from 25 to 60  eV over the frequency range from 120 to 156 GHz while at 226 GHz the LL width $\Gamma$  increases further until MIRO completely vanish at higher frequencies.  Such behavior needs to be understood and, it clearly cannot be explained by the existing theoretical models\cite{durst,ryzhii,vavilov,inarrea,dmitriev1,dmitriev2} for which the microwave radiation interacts directly with the conduction electrons since for these models the LL width does not depend on the MW frequency.  In fact, the theory with constant $\Gamma$  predicts that the number of observable oscillations should increase with increasing frequency because of the increasing density of states (Fig. 1(b)) in clear contradiction with the experiment, Fig. 1(a).  Let us discuss qualitatively the origin of this discrepancy and other unresolved experimental observations mentioned above.

\begin{figure}[h]
\includegraphics[width=8cm]{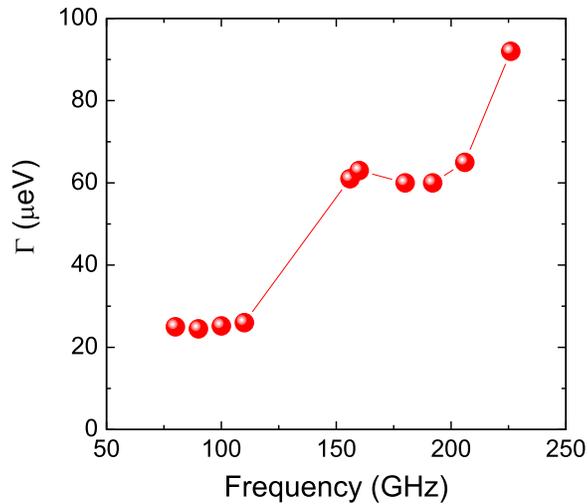}
\caption{(color online) The Landau level width $\Gamma$ as a function of microwave frequency determined from MIROs.  }
\end{figure}

\section{Discussion}
\subsection{Comparison between the two theoretical models.}
A key aspect of existing theoretical approaches is an assumption regarding the 2DEG microwave absorption process.  The two major theoretical models assume that MIRO result from single particle excitations between broadened LLs with a spatial displacement (scattering model) \cite{durst,ryzhii,vavilov} or directional diffusion due to the non-equilibrium distribution function  and modulated density of states.\cite{dmitriev1,dmitriev2}
\begin{figure}[h]
\includegraphics[width=8cm]{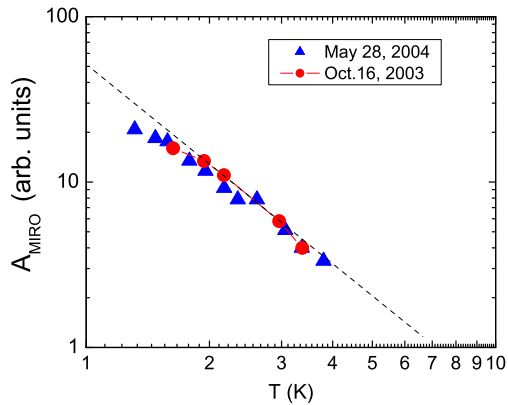}
\caption{(Color online) Temperature dependence of MIRO amplitude in a high-mobility GaAs/AlGaAs sample in arbitrary units determined by fitting experimental traces with eq. (1) using $\Gamma$  and $A$ as fitting parameters. Results of $\Gamma (T)$ dependence were published in Ref. \onlinecite{stud2}.}
\end{figure}
  Though these two models arrive at similar equations for the MIRO waveform, the non-equilibrium distribution function model predicts a roughly two orders of magnitude larger amplitude (eq. (17) in ref. \onlinecite{dmitriev2}) because in this model the non-equilibrium distribution function is accumulated during the diffusion process for $\Delta t \propto \tau_{in}$.
  Another difference between these two models is a different dependence of the pre-factor \emph{A(T)} in eq. (1) on temperature, which is much easier to test experimentally as compared to the frequency dependence.  The scattering model \cite{durst} predicts $A \propto 1/\tau_{tr}$ with $\tau_{tr}$ being the transport scattering time, which is a flat or very slow function of temperature at low temperatures.\cite{stud2}
  The non-equilibrium distribution function model predicts a much faster dependence of the amplitude $A \propto \tau_{in}$ (eq.(18) in ref.~\onlinecite{dmitriev1}), where  $\tau_{in}$ is the inelastic relaxation time.  At low temperatures in a degenerated 2DEG Fermi-liquid system the main mechanism for inelastic lifetime is associated with  the electron-electron interaction, which has the following dependence vs. temperature: $\tau_{in} \propto 1/(T^2\ln T) $    or roughly proportional to $T^{-2}$.\cite{giuliani}  In order to verify this prediction we have performed a temperature-dependent MIRO experiment.   Figure 4 shows dependence of MIRO amplitude determined by fitting experimental traces taken at the same MW power but different temperatures with eq. (1) using $\Gamma $  and $A$ as fitting parameters. It is evident from the figure that the observed dependence is very close to $1/T^2$ predicted by the distribution function model provided the inelastic scattering time at low temperatures is due to electron-electron interaction mechanism.\cite{dmitriev2}   Though this observation favors the distribution-function model, the final MIRO mechanism is still not resolved, because there are a number of other experimental observations which neither model can explain.
   In order to explain the experimental waveforms it is necessary to assume that the prefactor $A$ in eq. (1) does not depend on magnetic field an absorption strength of the single particle excitations is independent of  $\Delta N=N_f-N_i$, where $N_f$  and $N_i$ are the final and initial LL numbers.\cite{stud2}  On the other hand the direct absorption of light by harmonics of the cyclotron resonance is prohibited to first order and is usually very weak due to higher order corrections.  A common conjecture is that this absorption is activated by disorder, which contradicts the experimental observation that MIRO are observed only in samples with the highest possible mobilities meaning the least disorder.  It is clear, therefore, that the physics of MIRO is more complex, and in our opinion may be mediated by plasmons.  It should be noted that a similar mechanism of absolute negative photoconductivity was discussed in ref.~\onlinecite{ryzhii2}, but mediated by non-equilibrium optical phonons.

\subsection{Microwave magnetoabsorption.}
It was reported earlier in ref.~\onlinecite{stud3} that oscillation patterns are very different in MIRO transport and MW reflection experiments performed simultaneously.  No MIRO-like oscillations were observed in the MW reflection experiments but other kind of oscillations most likely due to 2D confined plasmons in a finite size sample.
At that time the plasmons structure was not clearly enough resolved to allow a comparison with theory due to the not well defined MW field distribution and the sample geometry which was additionally perturbed by contacts.
In order to improve the quality of the magneto-plasmon experiments we have performed the MW absorption measurements on a home-built tunable 40-60 GHz EPR system with a cylindrical cavity working in a TE011 mode\cite{seck1,seck2}.  In this experiment we used a better defined sample geometry without contacts.  The samples in this experiment were cleaved in a rectangular shape with a length L=7 mm and a width w=1.0 mm and w=0.5 mm, making the length of the sample much bigger than its width $L\gg W$.  The concentration of electrons was
1.9$\times$10$^{11}$ cm$^{-2}$ and the mobility 4.0$\times$10$^6$ cm$^2$/Vs.  In transport measurements these samples developed reasonably good MIRO similar to those in Ref.~\onlinecite{stud1}.

\begin{figure}[h]
\includegraphics[width=8cm]{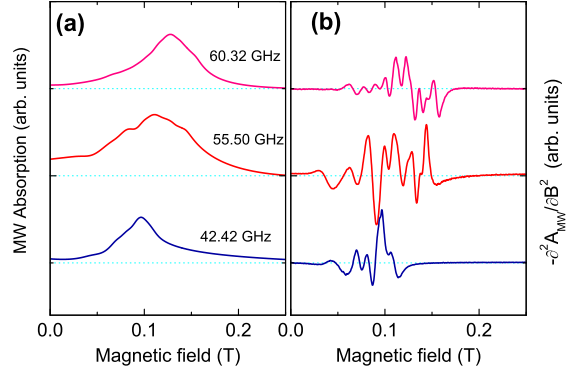}
\caption{(Color on line) \textbf{(a)} Examples of the microwave magneto-absorption spectra measured in EPR resonator on a rectangular 2DEG sample, w=1.0 mm;  \textbf{(b)} the second derivative of the corresponding spectra.}
\end{figure}

The position of the sample in the EPR resonator was chosen to be in the vicinity of the electric field maximum in such a way that the field was directed across the bar in the 2DEG plane.  Figure 5 shows an example of the magneto-absorption spectra (a) and its second derivative (b) for several frequencies. It is seen in panel (a) that the absorption of a finite size sample is a broad peak modulated with resonant structure due to confined magnetoplasmon excitations which are seen clearer in the second derivative.  Figure 6 plots the magnetic field position of the magneto-plasmon peaks for the two samples of two different widths (a) w=1.0 mm, and (b) w=0.5 mm.  The solid lines are calculated by a simple model for confined magnetoplasmon modes\cite{mikhailov2,volkov2}:
   \begin{equation}\label{2}
   f_p(B)=\sqrt{\omega_{0,N}^2+\omega_c^2}/2\pi,
\end{equation}
where $\omega_{0,N}=\sqrt{2\pi n_se^2q_N/m^*\kappa }  $  is the confined magneto-plasmon frequency in zero magnetic field, $n_s$ is the 2D electron density, $m^*$ is the effective electron mass, $\kappa $ is the dielectric constant (in our calculations we use  =12.8),
   $q_N\approx N\pi/W $ is the wave number of the $N$-th confined plasmon mode across the bar, $W$ is the width of the sample.  In Fig.~6~(a) there is a quite good agreement between  the experimental data and the simplified theoretical model by eq.~(2) (solid lines)  for the narrower sample in terms of the number of observable modes and their position vs. magnetic field for different frequencies. There are more modes observed for wider sample in Fig.~6~(b) in a qualitative agreement with eq.~(2).   
   However, the agreement for the wider sample is not as good due to many interfering peaks. The precision of the absorption measurements, in particular in the multi-mode regime on wider samples, is not very high since the plasmons modes are not very sharp. They have different oscillation strengths and therefore the weaker ones can be masked by other stronger peaks.\cite{mikhailov3}
\begin{figure}[h] 
\includegraphics[width=8cm]{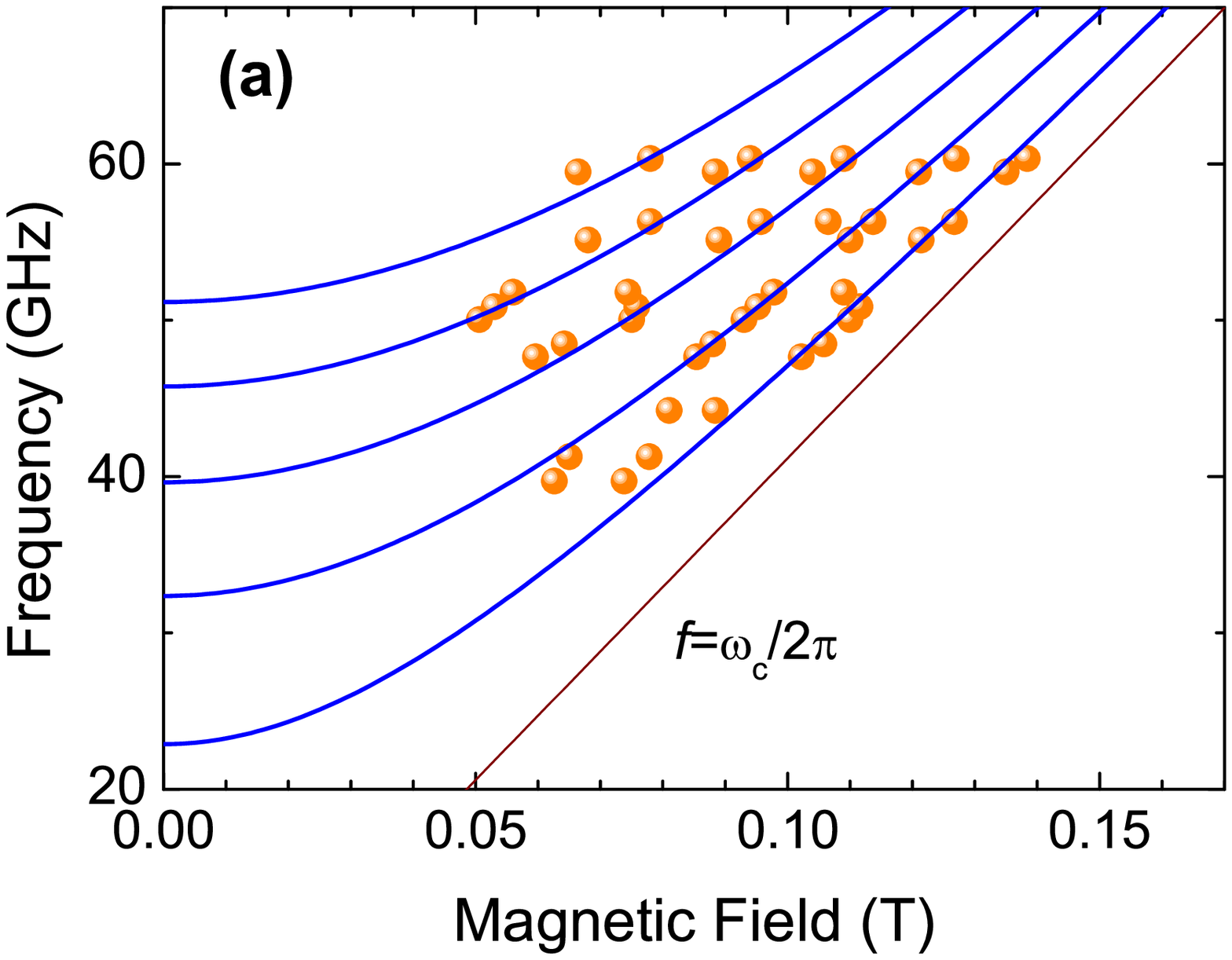}
\includegraphics[width=8cm]{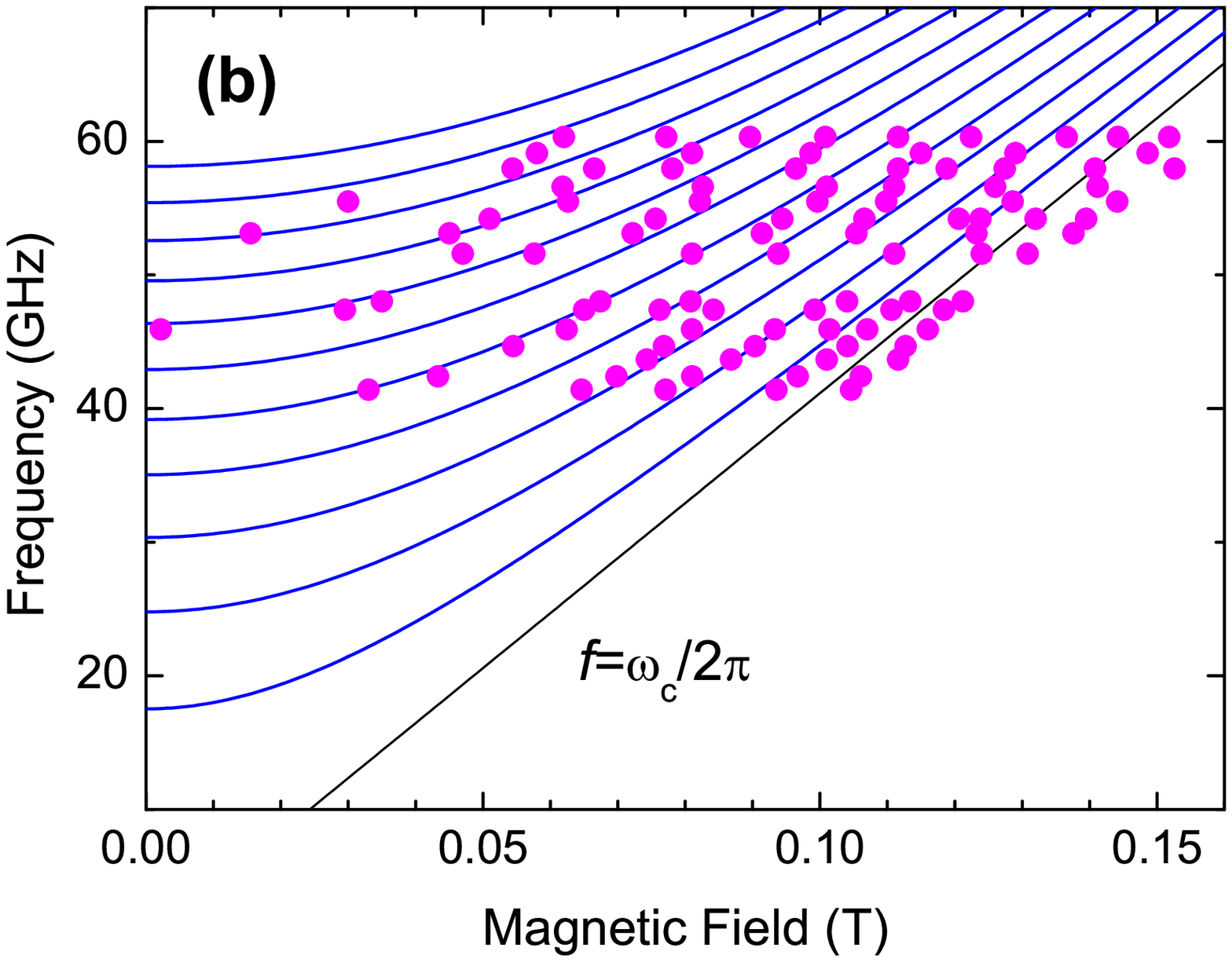}
\caption{(Color online) Frequency dependence of the  magnetoplasmon resonances vs. magnetic field for two GaAs/AlGaAs samples of two widths \textbf{(a)} w=0.5 mm, and \textbf{(b)} w=1.0 mm. }
\end{figure}
In addition, retardation effects\cite{kukushkin}  become more important in wider samples in Fig.~6~(B) which leads to a deviation of the plasmons dispersion from the simple theory and pushes the dispersion below the cyclotron resonance line in Fig.~6~(b). The detailed study of the magnetoplasmons in multimode regime in wide 2DEG stripes is outside of the scope of this work and will be published elsewhere.

 Therefore, it is evident from this experiment that many magneto-plasmon modes are excited in the MW experiments on rectangular samples in correspondence with the theoretical prediction.\cite{mikhailov3}
 Still, the above experiment cannot be considered as a direct, undisputed proof for magnetoplasmon involvement in MIRO.  However, in order to explain the unresolved experimental observations it is reasonable, at this stage, to assume that the microwave radiation induces plasmon oscillations, which then eventually have to decay.  One of the decay channels may be through excitation the single-particle inter-Landau level excitations, which in turn (consistent with the theoretical models) lead to the observation of MIRO.  Although, a fully consistent picture which integrates the physics of plasmon excitations with the appearance of MIRO still remains to be theoretically developed, it is clear that such an approach could indeed qualitatively explain many of the unexplained experimental facts.

\subsection{Qualitative explanation of the MIRO quenching via plasmons.}
We now turn to the principal effect reported in this paper - the MIRO quenching at higher MW frequencies.
The increase of the damping parameter, $\Gamma $ in Fig.~3 with frequency can be explained by involvement of plasmon oscillations as a result of the frequency dependent broadening of the plasmon excitations.  Broadening of the intermediate charge excitations may be viewed as an effective broadening of the excitation frequency, $\omega$  in eq.~(1).  Numerical simulations using eq.~(1) confirmed that a broadening of the excitation frequency can be considered as equivalent to a LL broadening.  Therefore, if the width of the plasmon excitations, $\Gamma_p$, increases with frequency\cite{ullrich}, the MIRO will experience an additional damping when $\Gamma_p$ becomes comparable to or larger than $\Gamma$.
Within this model it can be qualitatively explained why MIRO are not observed in optics or in the far infrared regime as being due to the plasmon lifetime decreasing at higher frequencies.   The number of excited plasmon modes increases quadratically with frequency as $N_p\approx (f/f_p)^2$, where $f_p$ is the fundamental plasmon mode for N=1 at B=0.  In our sample in Figs.(1)-(3) W$\approx $2 mm that gives us  $f_p\simeq$17 GHz and, for example, at \emph{f}=226 GHz we have $N_p\simeq$160.  Such large numbers of $N_p$ may lead to a decrease in the plasmon oscillator strength and, importantly, new decay mechanisms may occur for such high modes leading to the plasmon line broadening.

It is known that 2D plasmons have a very much reduced damping.  In particular, classical Landau damping is suppressed due to energy and momentum conservation laws.\cite{ullrich}
  In very high quality samples such as those used for MIRO experiments at low temperatures the lifetime of plasmon excitations, $\tau_p$, becomes very long (with $\Gamma_p=/\hbar\tau_p$), since the standard decay mechanisms are very much suppressed e.g. due to phonons and disorder.  Under these circumstances, other decay channels may become important through the inter-Landau level excitations, e.g. involving Bernstein modes.\cite{chaplik,lefebvre,krahne}
  In addition the resistance of 2DEG plays a selective role by being largely insensitive to absorption processes other than those responsible for MIRO. This is due to the highly degenerate nature of the 2DEG ($E_F>>k_BT$).
  In the same way, it is clear why the electron transition probability does not depend on the LL index change  $\Delta n$ and why MIRO do not depend on the circular polarization.\cite{stud2,smet} It is simply because the conduction electrons do not interact with microwaves directly but via charge oscillations so that the selection rules relevant for the direct interaction between light and free electrons do not apply.

\section{Conclusions}

In conclusion, we have experimentally studied MW magneto-absorption and the evolution of MIRO over a wide range of frequencies.  At MW frequencies below ~120 GHz the MIRO waveform is well described by a simple theoretical model, for which the Landau level width is the only fitting parameter.  The LL width extracted by this procedure remains constant for \emph{f}$\leq$120~GHz but rapidly increases for higher frequencies. The results can be qualitatively understood within a model for which the interaction between the microwave radiation and electrons is intermediated by plasmon excitations.  The excited magneto-plasmons may decay through the electron transitions between Landau levels giving rise to MIRO.

\begin{acknowledgments}
We thank V. Volkov, I. Dmitriev, and S. Dickmann for their interest and helpful discussions.
\end{acknowledgments}


\end{document}